\documentclass[epjCONF]{svjour}
\usepackage{graphicx}
\usepackage[varg]{txfonts} 
\usepackage[latin1]{inputenc}
\session-title{New technologies for probing the diversity of brown dwarfs and exoplanets}

\def\aj{AJ}%
\def\apj{ApJ}%
\def\apjl{ApJ}%
\def\aap{A\&A}%
\def\mnras{MNRAS}%
\def\sun{\hbox{$\odot$}}

\begin{document}
\title{Brown dwarfs forming in discs: where to look for them?}
\author{Dimitris Stamatellos\inst{1}\fnmsep\thanks{\email{D.Stamatellos@astro.cf.ac.uk}} \and Anthony Whitworth\inst{1} }
\institute{School of Physics \& Astronomy Cardiff University, 5 The Parade, Cardiff, CF24 3AA, UK }
\abstract{
A large fraction of the observed  brown dwarfs may form by gravitational fragmentation of unstable discs. This model reproduces the brown dwarf desert, and provides an explanation the existence of planetary-mass objects and for the binary properties of low-mass objects.  We have performed an ensemble of radiative hydrodynamic simulations and determined the statistical properties of the low-mass objects produced by gravitational fragmentation of discs. We suggest that there is a population of brown dwarfs loosely bound on wide orbits  ($100-5000$ AU) around Sun-like stars that surveys of brown dwarf companions should target. Our simulations also indicate that planetary-mass companions to Sun-like stars are unlikely to form by disc fragmentation.} 
\maketitle
\section{Introduction}
\label{intro}
The formation of brown dwarfs (BDs) is not well understood (\cite{whit2007}). Low-mass objects are difficult to form by gravitational fragmentation of unstable gas, as for masses in the BD regime ($\stackrel{<}{_\sim}~80~{\rm M}_{\rm J}$, where ${\rm M}_{\rm J}$ is the mass of Jupiter) a high density is required for the gas to be Jeans unstable. Thus, if BDs form according to the standard model of star formation, i.e. the collapse of a prestellar core,  then the pre-BD core must have a density of $\stackrel{>}{_\sim}10^{-16} {\rm g\ cm}^{-3}$. 

\cite{padoan} and \cite{hennebelle} suggest that these high density cores can be formed by colliding flows in a turbulent magnetic medium. However, this model requires a large amount of turbulence, and has difficulty explaining the binary properties of BDs. Additionally, the large number of brown-dwarf mass cores that the theory predicts  have not been observed.

Another way to reach the high densities required for the formation of BDs is in gravitationally unstable discs (\cite{whit06}, \cite{stam07b}). These discs form around newly born stars and grow quickly in mass by accreting material from the infalling envelope. They become unstable if the mass accreted onto them cannot efficiently redistribute its angular momentum outwards in order to accrete onto the central star (\cite{attwood09}).  This mechanism reproduces many of the observed statistical properties of  low-mass objects that are not satisfactorily explained by other theories, in particular the BD desert, the statistics of low-mass binary systems, and the formation of free-floating planets (\cite{stam09a},  \cite{stam09b},  \cite{stam09c}).  

BDs are also thought to form as ejected embryos from star forming regions, i.e. as by-product of the star formation process (\cite{rei}).  In this model BDs form the same way as low-mass stars, i.e. in collapsing molecular cores,  but shortly after their formation they are ejected from their parental core and stop accreting any further material. Hence, they do not realise their potential to become hydrogen-burning stars.

The above BD formation mechanisms are not exclusive of each other and can work in conjunction in star forming regions. For example, turbulent fragmentation can produce cores which collapse to form discs, the discs may then fragment to produce BDs, and the BDs may avoid accreting additional mass by being ejected.

In this paper we focus on the mechanism of BD formation by fragmentation of gravitationally unstable discs. We discuss the  properties of the BDs produced with this mechanism and  the predictions of the model that can be tested by observations.

\begin{figure*}
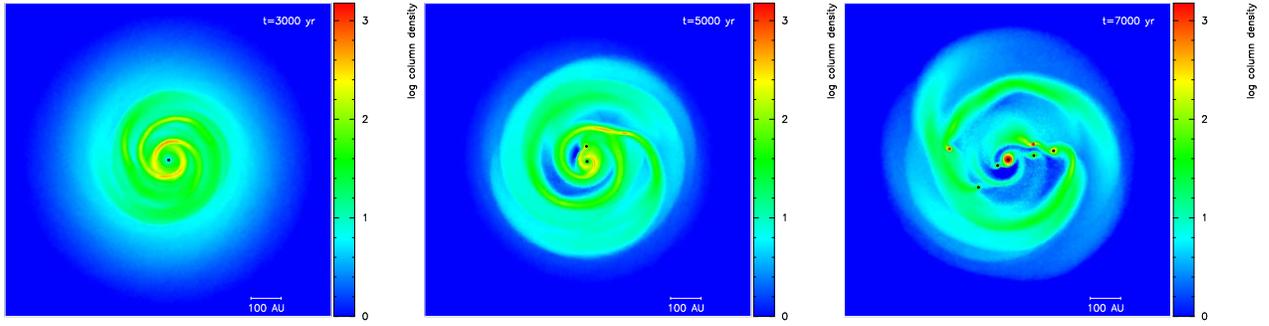

\centerline{
\includegraphics[angle=-90,width=5.5cm]{stamatellos_fig1.ps}
\includegraphics[angle=-90,width=5.5cm]{stamatellos_fig2.ps} 
\includegraphics[angle=-90,width=5.5cm]{stamatellos_fig3.ps}  
}
\caption{SPH radiative hydrodynamic simulation of the evolution of a $0.7\,{\rm M}_\odot$ disc around a $0.7\,{\rm M}_\odot$ star. The disc is represented using $2\times 10^6$ SPH particles. 3 snapshots are presented. The disc is gravitationally unstable and can cool efficiently, hence it quickly fragments to form low-mass H-burning stars, BDs, and planetary-mass objects. These objects form at radii from $\sim 100$ to $\sim 300\,{\rm AU}$ but due to mutual interactions a large fraction of  them escape  from the disc.}
\label{fig:sim}      
\end{figure*}

\section{Brown dwarf formation by fragmentation of gravitationally unstable discs}
\label{sec:1}

Discs can fragment if (i) they are massive enough that gravity overcomes the thermal and the local centrifugal support (\cite{toomre}), and (ii) they can cool fast enough  that the energy provided by the collapse of a proto-fragment is radiated away so that the growth of fragment continues (\cite{gammie}, \cite{rice}). Analytical and numerical studies have shown that the cooling time must be on the order of the dynamical time which happens to be similar to the orbital time. 

We have performed an ensemble of radiative hydrodynamical simulation of large unstable discs. The evolution of the disc is initially followed using Smoothed Particle Hydrodynamics (SPH), until $\sim 70\%$ of the disc mass has been accreted, either onto the stars condensing out of the disc, or onto the central primary star; this typically takes 10 to 20  kyr. Then the residual gas is ignored and the long term dynamical evolution of the resulting ensemble of stars is followed up to 200 kyr, using an N-body code.

The energy equation and associated radiative transfer are treated with the method of \cite{stam07a}. This method takes into account compressive heating or expansive cooling, viscous heating, radiative heating by the background, and radiative cooling.   The method performs well, in the optically-thin, intermediate and optically-thick regimes, and has been extensively tested (\cite{stam07a}). The gas in the disc is assumed to be a mixture of hydrogen and helium. We use an equation of state (\cite{black}, \cite{masunaga}) that accounts for the rotational and vibrational degrees of freedom of molecular hydrogen, and for the different chemical states of hydrogen and helium. We assume that ortho- and para-hydrogen are in equilibrium. For the dust and gas opacity we use the parameterization  by \cite{bell}, $\kappa(\rho,T)=\kappa_0\ \rho^a\ T^b\,$, where $\kappa_0$, $a$, $b$ are constants that depend on the species and the physical processes contributing to the opacity at each $\rho$ and $T$. The opacity changes due to ice mantle evaporation and the sublimation of dust are taken into account, along with the opacity contributions from molecules and H$^-$ ions.

We  assume a star-disc system in which the central  star  has initial mass $M_1=0.7\,{\rm M}_{\sun}$. Initially the disc has mass $M_{_{\rm D}}=0.7\,{\rm M}_{\sun}$, inner radius $R_{_{\rm IN}}=40\,{\rm AU}$, outer radius $R_{_{\rm OUT}}=400\,{\rm AU}$, surface density $\Sigma (R) \propto R^{-7/4}$, temperature $T(R)\propto R^{-1/2}$,  and hence approximately uniform initial Toomre parameter $Q\sim 0.9$. Thus, the disc is at the outset marginally gravitationally unstable. The radiation of the central star is taken into account.

Three snapshots from a typical simulation are shown in Fig.~\ref{fig:sim}. The gravitational instabilities develop quickly and the disc fragments within a few thousand years. Typically 5-10 objects form in each disc, most of them  BDs. The final stage of the system is determined by the subsequent accretion of material from the disc and mutual interactions that these objects undergo in the disc.

\section{The properties of objects formed by disc fragmentation}
\label{sec:2}

We have performed an ensemble of 12 radiative hydrodynamical simulation of unstable discs. These discs fragment on a dynamical timescale (a few thousand years) and $\sim 100$ objects are produced overall; mainly BDs but also hydrogen-burning stars and planetary-mass objects.

\subsection{The mass spectrum of the objects produced by disc fragmentation}

Most of the objects  ($\sim 70\%$) formed in the discs are BDs, including a few planetary-mass BDs. The rest are low-mass hydrogen-burning stars. The  typical mass of an object produced is $\sim 20-30~{\rm M}_{\rm J}$. The shape of the initial mass function (Fig.~\ref{fig:mspec}) is similar to what observations suggest, i.e. it is consistent with  $\Delta N/\Delta M\propto M^{-\alpha}$, where $\alpha\approx 0.6$ (e.g. in Pleiades  $\alpha=0.6\pm0.11$ {\cite{moraux03}, and in $\sigma$ Orionis $\alpha=0.6\pm0.1$ \cite{lodieu09}). 

\begin{figure}
\centerline{
\includegraphics[angle=-90,width=1\columnwidth]{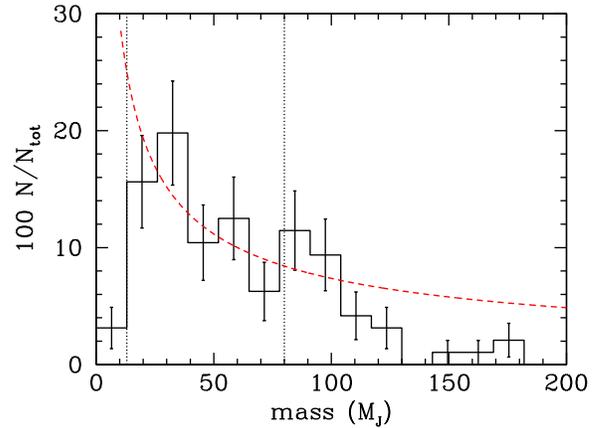}
}
\caption{The mass spectrum of the objects produced by disc fragmentation. Most of these are BDs ($70\%$); the rest are low-mass stars. The vertical dotted lines correspond to the D-burning limit ($\sim 13 M_{\rm J}$) and the H-burning limit ($\sim 80 M_{\rm J}$). The red dashed line refers to a low-mass IMF  with $\Delta N/\Delta M\propto M^{-0.6}$ (see text).}
\label{fig:mspec}      
\end{figure}

\subsection{Formation radius and the brown dwarf desert}

The discs fragment at distances from 100 to 300 AU from the central star. Most of the fragments form with an initial mass as low as 3 M$_{\rm J}$ (\cite{stam09c}). After a fragment forms it accretes mass from the disc, and interacts with the disc through drag forces and dynamically with other fragments. As a result some fragments migrate close to the central star. This region is rich in gas, hence these fragments accrete material from the disc and eventually become low-mass hydrogen-burning stars. Fragments that form farther out also accrete material from the disc but not quite as much; hence, they remain in the sub-stellar mass regime, i.e. they become BDs. If any of the BDs happen to migrate in the region close to the central star they tend to be ejected back into the outer region through 3-body interactions. Hence, the disc fragmentation model  produces a lack of BD close companions to Sun-like stars, i.e the BD desert (Fig.~\ref{fig:desert}). The BD desert may extend out to 300 AU but it is less dry outside 100 AU. Most of the BD companions formed by disc fragmentation have eccentric orbits, hence they could spend some time within 100 AU from the central star  during their periastron passage. 

\begin{figure}
\centerline{
\includegraphics[angle=-90,width=1\columnwidth]{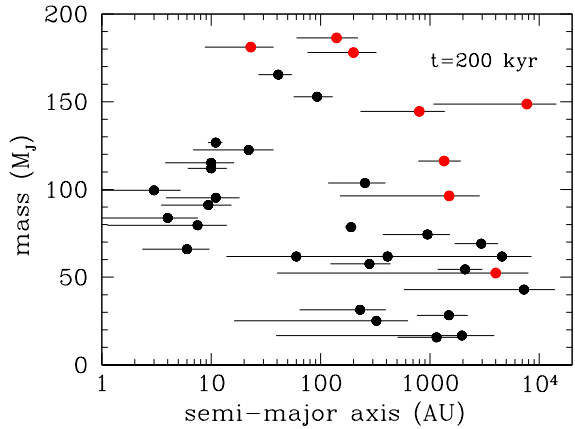}
}
\caption{The BD desert. The orbital semi-major axes of the objects  formed by disc fragmentation and remain bound to the central star plotted against the mass of the objects. The bars indicate the minimum and maximum extent of the orbit. There is a lack of BD companions close to central star, but there is a population of BDs loosely bound to the central star (at distances $\sim 100-5000$ AU).}
\label{fig:desert}      
\end{figure}

\subsection{Where to look for brown dwarfs forming in discs?}

A large fraction  of the BDs forming in discs are ejected into the field, with typical  velocities of $\sim1-3\ {\rm km\ s^{-1}}$. The BDs that remain bound to the central star are on wide orbits, i.e. they are loosely bound. Hence, we predict a population of BDs at distances from 20 to 5000 AU from the central star. This is the region that should be targeted by surveys looking for BD companions to Sun-like stars. However, in a cluster environment such loosely bound companions could be liberated (\cite{goodwin07}) and become field BDs; hence, the number of BDs on wide orbits should be smaller than what our simulations of isolated discs predict.

\subsection{Brown dwarf discs}

Most of the BDs form with discs with masses of a few tens of ${\rm M}_{\rm J}$  and radii of a few tens of AU. These are consistent with observations of BD discs (\cite{rayjay03}, \cite{luhman05}). 

\subsection{Free-floating planetary-mass objects}

The disc fragmentation model provides an explanation of the existence of free-floating planetary-mass objects (\cite{lucas00}, \cite{zapatero00}). In our model these objects form in the disc by gravitational fragmentation and quickly after their formation they are ejected from the system due to 3-body interactions. Hence, they stop accreting and their mass remains low. Based on the specific initial conditions we examine, we predict that free-floating BDs outnumber free-floating planetary-mass objects by a factor of at least 10.

\subsection{Low-mass binary statistics}

Close and wide BD-BD binaries are a  common outcome of disc fragmentation. The components of the binary form independently in the disc and then pair up. The simulations produce  all kinds of low-mass binaries:  star-star, star-BD , BD-BD, and BD-planetary mass binaries. 

 \paragraph{Binary fraction.} Our model predicts a low-mass binary fraction of $16\%$. This is comparable with the low-mass binary fraction observed in star-forming regions (e.g. in Taurus : $> 20\%$ \cite{kraus06}, in the field $15\pm5 \%$, \cite{gizis03}). However, these are optical surveys and they are not able to detect tight binaries with separation of a few AU. Very tight binaries can be probed by  radial velocity surveys (e.g. \cite{joergens08}), which find  a low-mass tight-binary fraction of $10-30\%$. This is roughly consistent with our model. However, our model produces either very tight  or very wide binaries (\cite{stam09a}).

\paragraph{Wide and close low-mass binaries.} We predict that close low-mass binaries should outnumber wide ones, and this seems to be what is observed (\cite{burgasser07}).  A few of the low-mass binaries, including one of the very wide systems, have become unbound from the central star. Therefore the ejection mechanism does not militate against delivering wide low-mass binaries to the field. Amongst the low-mass binaries that remain bound to the central star, the semi-major axis $a$ of the orbit around the central star tends to increase as the total mass of the low-mass binary decreases. 

\paragraph{Mass-ratios of low-mass binaries.} Most of the low-mass binaries ($55\%$) have components with similar masses ($q>0.7$), in agreement with the observed properties of low-mass binaries (e.g. \cite{burgasser07}).

\paragraph{Low-mass binaries as companions to Sun-like stars.} The model also predicts that BDs that are companions to Sun-like stars are more likely to be in binaries (binary frequency 25\%) than BDs in the field (frequency $5\;{\rm to}\;8\%$). This trend is comparable to what is observed, although the observed binary frequencies are somewhat higher. \cite{burgasser05} report a binary fraction of $45^{+15}_{-13}\%$ for BD companions to Sun-like stars, and a binary fraction of only $18^{+7}_{-4}\%$ for BDs in the field. This is corroborated by the results of \cite{faherty09}.

\section{Can planetary-mass objects form by disc fragmentation?}

A large number of exoplanets has been observed around Sun-like stars. It has been suggested that these planets could form by gravitational fragmentation of discs (\cite{boss00}, \cite{boley09a}). In our simulations most of the fragments initially have masses as low as 3~M$_{\rm J}$ (\cite{stam09c}, see also \cite{boley09b}). These fragments form in the outer disc region ($>100$~AU) as the inner disc region cannot cool fast enough (\cite{stam08}). However, these objects quickly grow in mass by accreting material from the disc. Only fragments that are quickly ejected from the disc after their formation end up with masses below 13 M$_{\rm J}$. Hence, it is unlikely that the exoplanets observed close  to Sun-like stars form by gravitational fragmentation of discs.

\section{Observing fragmenting discs}

Typically a few BDs form in each fragmenting disc around a  Sun-like star. Hence, it may be that only  $\sim20\%$ of Sun-like stars need to have large and massive unstable discs to produce a large fraction of the observed BDs.  Assuming that the lifetime of the Class 0 phase is $10^5$ yr and that in the disc fragmentation scenario the disc fragments and therefore dissipates within $10^4$ yr , then the probability of observing a fragmenting disc around a Class 0 object is only 10\%. Then, considering that only  20\% of Sun-like stars may have such unstable discs, the probability of observing such discs is only 2\%. Hence, fragmenting discs should be very difficult to find (\cite{maury09}), but there are hints of their existence (\cite{andrews09}).

\section{Conclusion}

The mechanism of BD formation by fragmentation of gravitationally unstable discs reproduces the BD desert around Sun-like stars, and can explain the existence of free-floating planetary mass objects and the binary properties of low-mass objects. We suggest that there is a population of BDs on wide orbits around Sun-like stars that could be probed with future observations.

\end{document}